\documentclass[twocolumn,showpacs,preprintnumbers,amsmath,amssymb]{revtex4}
\usepackage{graphicx,epsf}
\begin{document}

\title{Scale Invariant Fractal and Slow Dynamics in Nucleation and Growth Processes}

\author{M. Kamrul Hassan$^{\dagger}$ and J\"urgen Kurths$^\ddagger$}
\affiliation{
$^\dagger$ University of Dhaka, Department of Physics, Theoretical Physics Division, Dhaka 1000, Bangladesh \\ 
$^\ddagger$ University of Potsdam, Department of Physics, Postfach 601553, D-14415 Potsdam, Germany 
}

\begin{abstract}%
We propose a stochastic counterpart of the classical Kolmogorov-Johnson-Mehl-Avrami (KJMA) model to describe the nucleation-and-growth phenomena of a stable phase (S-phase). We report that for growth velocity of S-phase $v=s(t)/t$ where $s(t)$ is the mean value of the interval size $x$ of metastable phase (M-phase) and for $v=x/\tau(x)$ where $\tau(x)$ is the mean nucleation time, the system exhibits a power law decay of M-phase. We also find that the resulting structure exhibits self-similarity and can be best described as a fractal. Interestingly, the fractal dimension $d_f$ helps generalising the exponent $(1+d_f)$ of the power-law decay. However, when either $v=v_0$ (constant) or $v=\sigma/t$ ($\sigma$ is a constant) the decay is exponential and it is accompanied by the violation of scaling.

\end{abstract}


 \maketitle

The kinetics of phase transformation via nucleation and growth of a stable phase (S-phase)
represents one of the most fundamental topic of interest in both science and technology \cite{kn.christian}. It plays a key role in metallurgical applications as well as in many seemingly unrelated fields of research.  
The kinetics of crystal growth \cite{kn.crystallization}, the domain switching phenomena in 
ferroelectrics \cite{kn.ferroelectrics} and the DNA replication in organisms \cite{kn.replication} are just a few examples to name. Much of our theoretical understanding of such phenomenon is provided by the Kolmogorov-Johnson-Mehl-Avrami (KJMA) model. It has been formulated independently by Kolmogorov, by Johnson and Mehl and by Avrami  
in and around the $1940$s \cite{kn.kolmogorov}. 
It still remains one of the most studied theory and is the only means of interpreting the experimental data to gain insight of the process. 
The basic algorithm of the KJMA theory is trivially simple. One time unit of the process in one dimension can be defined as follows. 
\begin{itemize}
\item[{\it i)}] Randomly a position is selected for nucleation of point-like seed of S-phase
on an interval of M-phase.
\item[{\it ii)}] Upon nucleation the seed starts growing isotropically on either side with a constant velocity $v_0$ at the expense of the decay of M-phase.
\item[{\it iii)}] Whenever two such growing phases from opposite sides meet, growth ceases immediately at the point of contact, while continuing elesewhere. 
\item[{\it iv)}] The process is repeated {\it ad infinitum}. 
\end{itemize} 

In the context of nucleation and growth phenomena, the question we usually ask is: What is 
the fraction of M-phase that still survives at time $t$? According to the KJMA theory, such quantity in $d$ dimensions obey the exponential decay known as the Kolmogorov-Avrami law
\begin{equation}
\label{eq:k-a}
\Phi(t)=\exp\Big [-{{\Omega_d}\over{d+1}}\Gamma v_0^d t^{d+1}\Big ],
\end{equation}
where, $\Gamma$ decribes the constant nucleation rate per unit volume and $\Omega_d$ is the constant volume factor of the $d$ dimensional hypersphere
($\Omega_d =1,\pi,4\pi/3$  for $ d=1,2,3$ respectively and $d+1$ is often called as the Avrami exponent) 
\cite{kn.kolmogorov}. Recently, the derivation of 
correlation function and its connection to the scattering cross-section \cite{kn.sekimoto}, 
theory of grain-size populations \cite{kn.population}, and the generalization of the KJMA theory to multiple stable phase \cite{kn.multiphase} has made a significant contribution as it now provide a better means of interpreting the experimental data.
Alongside, there have been reports too that the experimental data in 
some cases do not fit a straight line 
in the plots of $\log\log[\Phi(t)]$ against $\log[t]$ \cite{kn.price,kn.sessa}. It clearly raises some concerns and hence it requires further theoretical interest in order to find varients of the model which would be suitable under different physical situations. 

In this paper, we lift the restriction imposed by the rule {\it (ii)}, the conventional constant growth velocity, by choosing various form of velocities. Precisely, we choose velocities that decreases as the system proceeds and check how such decreasing grwoth velocities of different strength influence the resulting dynamics. 
In fact, a more realistic theory does require the inclusion of a growth velocity that 
may change in the course of time. The idea to move away from the constant growth velocity is born by the following observation. The problem in question is intrinsically a kinetic one and hence, as the time proceeds, the size and shape of both the phases change continuously altering the strength of their competition between the surface tension and the free energy density difference between the two phases. The strength of this competition is the one that ultimately determines the growth velocity. This implies that the growth velocity of S-phase should therefore depend either on time or on the size and shape of one, or both, of the phases or on proper combinations of all these parameters in some sense. 
The inlcusion of velocity, that is not constant, appears to reveal a number of new interesting results. The most significant of all is the emergence of the scale invariant fractal geometry accompanied by a power-law decay of  M-phase instead of the traditional exponential decay. We show that the system exhibits such fascinating behaviour only if the growth velocity decreases as the system eveolves. To be more precise, the systems exhibit scale-free fractal if the growth velocity at any given time $t$ depends on the mean size $s(t)$ of M-phase i.e., 
$v=s(t)/t$ or if it depends on the mean nucleation time $\tau(x)$ i.e., $v=x/\tau(x)$.

In this work, we mainly focus on extracting the key features of the phenomena (at least thier qualitative behaviour) with the least 
possible mathematical complexity. We therefore choose the one-dimensional (1d) system as we can solve it for some special cases analytically and exactly. 
The 1d model can in fact fully capture the qualitative bahaviour of the key quantities of interest namely the decay of M-phase. The Kolmogorov-Avrami formula itself is a testament to its justification which clearly shows that the exponential decay of M-phase is common to all dimensions.
The distribution function $C(x,t)$ of M-phase of size $x$ at time $t$ for random nucleation and growth mechanism evolves according to the following integro-differential equation
\begin{eqnarray}
\label{eq:g}
{{\partial C(x,t)}\over{\partial t}}& = & -xC(x,t)+2\int_x^\infty C(y,t)dy \\ \nonumber & + & {{\partial}\over{\partial x}}\big[v(x,t)C(x,t)\big].
\end{eqnarray}
The first two terms on the right hand side represent the decrease and increase of interval size $x$ of M-phase due to the nucleation of S-phase on $x$ and $y>x$ respectively. Every nucleation event creates two new intervals. The factor $`2'$ in the integral term implies that either 
of the two new intervals can be of size $x$. Finally, the third term on the right hand side
is derived by taking into account all the possible ways the distribution function $C(x,t)$ still remains in the size range $[x,x+dx]$ despite loss and gain of $x$ due to continuous growth in a span of infinitesimal time $dt$. For instance, if we assume that the process is homogeneous in time then all the possibilities are 
\begin{eqnarray}
\label{eq:3}
\left. C(x,t+dt) \right |_{{\rm growth}} & = & \big [1-v(x,t)dt/dx \big ]C(x,t) + 
\\ \nonumber &  &  \big [v(x+dx,t)dt/dx \big ]C(x+dx,t),
\end{eqnarray}
which is in fact the third term on the right of Eq. (\ref{eq:g}). Here, $v(x,t)dt/dx$ ($v(x+dx,t)dt/dx$) is the fraction of concentration $C(x,t)$ that is lost (gained) in time $dt$ due to the growth of  S-phase.

We now attempt to solve Eq. (\ref{eq:g}) subject to the initial conditions 
\begin{equation}
\label{eq:initial}
C(x,0)={{1}\over{L}}\delta(x-L), \hspace{0.6cm} 
\lim_{L \longrightarrow \infty}\int_0^\infty C(x,0)dx =0.
\end{equation}
This ensures that there is no seed of S-phase at $t=0$. 
Once we know $C(x,t)$, we can immediately find the fraction of  M-phase that 
remained untransformed $\Phi(t)=\int_0^\infty xC(x,t)dx$ and the number 
density of the intervals of M-phase $N(t)=\int_0^\infty C(x,t)dx$. 
The fraction of  M-phase covered by S-phase is 
related to $\Phi(t)$ via $\theta(t)=1-\Phi(t)$ that evolves as
\begin{equation}
 \label{eq:10}
{{d\theta(t)}\over{dt}}=\int_0^\infty v(x,t)C(x,t)dx.
\end{equation}
The quantities $\Phi(t)$ and $N(t)$ can also 
be used to know how the average interval size $s(t)=\Phi(t)/N(t)$ varies in time. 

The two parameters involved in the definition of velocity are length and time. The growth velocities that we can choose are as follows: (i) $v(x,t)=v_0$, the traditional constant growth velocity (classical KJMA model), (ii) $v(x,t)=\sigma/t$ where $\sigma$ is a constant and bears the dimension of length, (iii) $v(x,t)=s(t)/t$ where $s(t)$ is the mean interval size of M-phase at time $t$ and finally (iv) $v(x,t)=m x/\tau(x)$ where $\tau(x)$ is the mean nucleation time and $m$ is a dimensionless positive constant \cite{kn.maslov}.

For growth velocities {\it (a)}\ -  {\it (c)}, which are 
independent of $x$, the solution of Eq. \ (\ref{eq:g}) can be obtained by 
substituting the ansatz
\begin{equation}
\label{eq:7}
C(x,t)=B(t)\exp[-xt].
\end{equation}
Here the time dependent prefactor $B(t)$ obeys the ordinary differential equation 
\begin{equation}
\label{eq:8}
{{d\ln B(t)}\over{dt}}=2/t-v(t)t.
\end{equation}
We have to solve Eq. (\ref{eq:8}) subject to the initial condition $B(0)=0$, followed by  Eq. 
(\ref{eq:initial}). To proceed further and for clarity, we will treat each case independently.

{\it (a)} Solving Eq. (\ref{eq:8}) for $v=v_0$  we obtain 
$B(t)=t^2\exp[-v_0t^2/2]$ and therefore
\begin{equation}
 \label{eq:9}
C(x,t)=t^2\exp[-xt-v_0t^2/2].
\end{equation} 
Using this in the definition of $\Phi(t)$, we immediately find $\Phi(t)=\exp[-v_0t^2/2]$
which is the celebrated Kolmogorov-Avrami formula in one dimension, except the factor $\Gamma$.
Note that Eq. (\ref{eq:g}) describes the random 
sequential nucleation of one seed at each time step and hence $\Gamma=1$. 
It clearly demonstrates not only how well the theory works but also its simplicity. 
Furthermore, we find that the number density of intervals of M-phase 
changes as $N(t)\sim te^{-v_0t^2/2}$. It reveals that $N(t)$ 
rises linearly at the early stage due to the nucleation of  S-phase. At the late stage on the other hand, $N(t)$ decreases exponentially due to the fast coalescence of the neighbouring stable phases \cite{kn.naim}. The mean interval size $s(t)$ on the other hand decays as
 $s(t)\sim t^{-1}$. 

{\it (b)} We now solve Eq. (\ref{eq:8}) for the velocity $v(t)=\sigma/t$. 
to give
\begin{equation}
C(x,t)=t^2\exp[-(x+\sigma)t].
\end{equation} 
In this case too, we find that the fraction of  
M-phase decays exponentailly $\Phi(t)=\exp[-\sigma t]$ but slower than that for constant 
velocity.  
Here the Avrami exponent $\alpha=1$ which corresponds to the one typically known for the  heterogeneous nucleation and growth processes \cite{kn.heterogeneous1} despite the fact in the present case it strictly describes the homogeneous nucleation.
It proves that the Avrami exponent not only depends on the nature of the nucleation process but it also depends on the nature of the growth velocity. We also find that 
the number density of  M-phase varies as $N(t)\sim te^{-\sigma t}$. This reveals that at the early 
stage, it rises linearly like the model {\it (a)}. However, at the late stage, the coalescence takes place less frequently than in the model {\it (a)}.
 
{\it (c)} In order to be able to solve Eq. (\ref{eq:g}) for $v(x,t)=s(t)/t$, we first need to know explicitly how the mean interval size $s(t)$ decays in time. We have already solved Eq. (\ref{eq:g}) for two different growth velocities and in both cases we found that the mean interval $s(t)$ decreases with time as $s(t)\sim t^{-1}$. Interestingly. it is also the case without the growth term which is known as random scission model \cite{kn.ziff}. All this suggests that $s(t)= kt^{-1}$ is the most generic choice for the mean interval size, where $k$ is a dimensionless parameter. Incorporating it into the defintion {\it (c)} gives 
$v(x,t)=k/t^2$. Solving Eq. (\ref{eq:8}) for this growth velocity yields
\begin{equation}
\label{eq:c}
C(x,t)=t^{2-k}\exp[-xt].
\end{equation}
Unlike the previous two cases, it is of particular interest for the following reasons. We 
find that all the moments, $M_n(t)$
of $C(x,t)$ where $M_n(t)=\int_0^\infty x^nC(x,t)dx$, exhibit a power-law behaviour 
$M_n(t) \sim t^{-(n-(1-k))}$ and therefore, 
\begin{equation}
\Phi(t)=t^{-k}.
\end{equation}
The emergence of such a slow decay makes it a good candidate to look for a scale invariance in the system. The $k$ value here plays 
the crucial role in determining the dynamics of the system and is bound by the lower and the upper limits and they are fixed by the following physical constraints. The lower bound is fixed by the behaviour of $\Phi(t)$ that should be increasing function of time and hence it demands $k>0$. The upper bound on the other hand is fixed by the constraint that the number density $N\sim t^{1-k}$ should be an increasing function of time, at least, during the early and intermediate stage. This immediately provides the upper bound $k<1$. The only non-trivial and physically interesting $k$ value are the ones that stay within the interval $[0,1]$. In passing, we note that in this model too the average interval size $s(t)$ decays as $s(t)=t^{-1}$. 

{\it (d)} 
Finally, we consider the case where $v(x,t)= mx/\tau(x)$, however, according to Eq.(\ref{eq:g}), the typical time 
$\tau$ between two nucleation events on an interval 
of size $x$ is $\tau=x^{-1}$ and therefore $v(x,t)=mx^2$ revealing that the growth velocity 
decreases increasingly fast as the size of M-phase decreases since $x$ itself is a decreasing
quantity with time. For simplicity reason, we only consider the case $m=1$. 
Invoking a simple dimensional analysis allows us the following scaling ansatz
\begin{equation}
\label{eq:ansatz}
C(x,t)\sim t^w\phi(xt),
\end{equation}
such that $t^w$ bears the dimension of $C(x,t)$ and $\phi(\xi)$ is the scaling function. 
Upon substituting the ansatz into Eq. (\ref{eq:g}) and seeking the self-similar
solution immediately reduces the original partial integro-differential equation into an ordinary 
hypergeometric differential equation for $\phi(\xi)$
\begin{equation}
\xi(1-\xi){{d^2\phi(\xi)}\over{d\xi^2}} 
+\{(1+w)-3\xi\}{{d\phi(\xi)}\over{d\xi}}+\phi(\xi)=0.
\end{equation} 
The most nontrivial feature of $\phi(\xi)$ is its universality, in the sense that it is independent 
of the initial condition. The only physically acceptable solution of the above equation is 
\begin{equation}
\phi(\xi)= ~_2F_1(1+\sqrt{2},1-\sqrt{2};1+w;\xi),
\end{equation}
and asymptotically it behaves as $\phi(\xi)\sim \exp[-(\sqrt{2}-1)\xi]$ \cite{kn.luke}. 
The yet undetermined exponent $w$ of the scaling ansatz is fixed by the conservation principle. However, 
as the intervals of  M-phase are lost to  S-phase, the system violates the simple mass conservation. 
Nevertheless, it still obeys a conservation law although this time it is a non-trivial quantity as we 
shall see below. To do so, we 
incorporate the definition of $M_n(t)$ into Eq. (\ref{eq:g}) and then, after some simple algebraic 
manipulations, we are able to write the following rate equation for $M_n(t)$
\begin{equation}
{{dM_n(t)}\over{dt}}=-{{(n-\alpha)(n+\alpha+2)}\over{(n+1)}}M_{n+1}(t).
\end{equation}   
Thereafter, demanding the steady state 
condition $dM_n(t)/dt=0$ immediately reveals that it is the value $\alpha=(\sqrt{2}-1)$ that fixes the 
exponent $w=\sqrt{2}$. The asymptotic solution for $C(x,t)$ is therefore,
\begin{equation}
C(x,t)\sim t^{\sqrt{2}}\exp[-(\sqrt{2}-1)xt].
\end{equation} 
We thus find that $\Phi(t)$ decays algebraically with a non-trivial exponent $0.5857$ whereas $N(t)$ 
grows with an exponent $0.4142$. 
However, the average interval of M-phase $s(t)$ decays exactly like all previous three cases
i.e., $s(t) \sim t^{-1}$.
For model {\it (c)} too, we could substitute the ansatz Eq. (\ref{eq:ansatz})
into Eq. (\ref{eq:g}) and find that the scaling function $\phi(\xi)$ obeys  
an ordinary differential equation. Its solution would have been the
same as we obtained by using the direct method, Eq. (\ref{eq:c}). This further supports the fact that the model {\it (c)} admits scaling. 
However, applying the same treatment to the model 
{\it (a)} and {\it (b)} 
would immediately reveal that neither of these two cases yield a 
reduction to an ordinary differential equation of one variable. This suggests that the  model {\it (a)} and {\it (b)} do not admit a scaling solution.
In fact, the exponential decay of  M-phase, 
is a clear signature of the violation of scaling. 
It is note worthy to mention that the number density in {\it (c)} and {\it (d)} increases for all 
time - a sharp contrast with the models {\it (a)} and {\it (b)} where 
$N(t)$ increases only at the early stage. This is due to the fact that in models {\it (c)} and {\it (d)}, 
the growth velocity decreases in time in such a way that two growing phases from opposite direction 
hardly coalesce. In fact, the growth of  S-phase virtually stops prior to coalescence. 

The existence of scaling and the non-trivial exponent of $\Phi(t)$ provides an extra motivation to go 
beyond the simple scaling description. This can be done by invoking the idea of fractal analysis, as it has been a very 
useful tool to obtain a global exponent called fractal dimension.  
To do so, we need a proper yardstick to measure the size of the set created in the 
long time limit. The most convenient one is the average interval size $s(t)\sim t^{-1}$. Using 
this yardstick, we find that the number of $s(t)$ we need to cover the system scales as
\begin{equation}
\lim_{s(t) \longrightarrow 0}N(s(t))\sim s(t)^{-d_f}.
\end{equation}
The exponent $d_f$ is known as the fractal dimension or the Hausdorff-Besicovitch dimension 
\cite{kn.fractal} of the resulting set created by the nucleation and growth process. For 
$v(x,t)=k/t^2$, we find $d_f=(1-k)$ and $v(x,t)=x^2$ yields $d_f=0.4142$. For general $m$ in
{\it (d)}, the fractal dimension is $d_f(m)=-{{1}\over{2}}(1+1/m)+{{1}\over{2}}\sqrt{(1+1/m)^2+4/m}$ (for further details see \cite{kn.hassan}). In both cases ({\it (c)} and {\it (d)}), the Kolmogorov-Avrami formula can be replaced by the following general power-law decay of  M-phase
\begin{equation}
\label{eq:final}
\Phi(t)\sim t^{-(1-d_f)}.
\end{equation}
It reveals a generalised exponent $(1-d_f)$ that can quantify the extent of decay of M-phase. The 
coverage $\theta(t)$ by the S-phase reaches its asymptotic value $\theta(\infty)=1$ again follow a power-law
\begin{equation}
\theta(\infty)-\theta(t) \sim t^{-(1-d_f)},
\end{equation}
which is reminscent of the Feder's law in RSA  \cite{kn.feder}.

To put it into a context, we have studied the random nucleation and growth processes of a stable 
phase for different growth velocities. So far, the KJMA model has only been been studied with constant velocity. 
However, we conceive in systems where the dynamics is governed by the nucleation and 
growth processes, it is more logical to consider that the growth velocity of S-phase may depend on the size
of M-phase present at every instant of time. Invoking the idea of such velocites, 
we have found the following interesting and previously unknown results. The average interval size $s(t)$ of M-phase 
shows a power-law decay $s(t)\sim t^{-1}$ irrespective of the detailed choice of the velocity. The fraction of M-phase that 
still survive at any given time is very sensitive to the specific choice of the velocity. The 
resulting structure exhibits a scale-invariant fractal if only the growth velocity depends 
on either the mean size of the M-phase $s(t)$ or on the mean nucleation time $\tau(x)$ at every instant of time. 
Moreover, only in such cases the system show a power-law decay of the M-phase and we hve shown that exponent 
of the decay law can be generalised $(1-d_f)$ by incorporating the fractal dimension $d_f$. 
However, in the case where both or one of the parameter of the definition of velocity is assumed a constant value 
(e.g., $v=v_0$ or $v={\rm constant}/t$) the decay is exponential in nature which is always accompanied by 
the violation of scaling. 

We conclude with the following words. The power-law decay of  M-phase can be assumed to be a generalized formula replacing the classical Kolmogorov-Avrami law, provided, the distribution of  M-phase in the late stage describes a scale-free fractal. The fractal diemsnion $d_f$ is the quantitative measure of the notion that the density of  M-phase is less at larger length scale. We believe that the present work will have a significant impact in changing the way we 
intended to interpret the experimental data as we are now aware of the fascinating results due to the decelerating growth velocity.

\noindent
One of us (M. K. H.) acknowledges the financial support from the Alexander von Humboldt 
foundation.

\end{document}